\documentclass[conference]{IEEEtran}
\IEEEoverridecommandlockouts
\usepackage{cite}
\usepackage{amsmath,amssymb,amsfonts}
\usepackage{algorithmic}
\usepackage{graphicx}
\usepackage{textcomp}
\usepackage{url}
\usepackage{xcolor}
\usepackage{listings}
\usepackage[T1]{fontenc}
\def\BibTeX{{\rm B\kern-.05em{\sc i\kern-.025em b}\kern-.08em
    T\kern-.1667em\lower.7ex\hbox{E}\kern-.125emX}}
\begin{document}

\title{Performance Analysis of Quantum-Secure Digital Signature Algorithms in Blockchain}

\author{
    \IEEEauthorblockN{Tushar Jain}
    \IEEEauthorblockA{
        \textit{Institute of Computer Science} \\
        \textit{University of Tartu} \\
        Tartu, Estonia \\
        tushar.jain@ut.ee}
}

\maketitle

\begin{abstract}
The long-term security of public blockchains strictly depends on the hardness assumptions of the underlying digital signature schemes. In the current scenario, most deployed cryptocurrencies and blockchain platforms rely on elliptic-curve cryptography, which is vulnerable to quantum attacks due to Shor's algorithm. Therefore, it is important to understand how post-quantum (PQ) digital signatures behave when integrated into real blockchain systems. This report presents a blockchain prototype that supports multiple quantum-secure signature algorithms, focusing on CRYSTALS-Dilithium, Falcon and Hawk as lattice-based schemes. This report also describes the design of the prototype and discusses the performance metrics, which include key generation, signing, verification times, key sizes and signature sizes. This report covers the problem, background, and experimental methodology, also providing a detailed comparison of quantum-secure signatures in a blockchain context and extending the analysis to schemes such as HAETAE.
\end{abstract}


\section{Introduction}
Blockchains are a distributed list of transactions maintained by a network of untrusted nodes \cite{Nofer2017,8024092,buterin2013ethereum}. Between these nodes, the integrity and authenticity of transactions are maintained through digital signatures. Every transaction is signed by the owner of the private key, and nodes validate signatures before accepting new blocks. In deployed systems such as Bitcoin and Ethereum, these signatures are typically produced with elliptic-curve cryptography (ECC), for example, ECDSA or Schnorr signatures over secp256k1 \cite{buterin2013ethereum,inbook}. The security of these schemes is dependent on problems such as the discrete logarithm, which are believed to be safe against classical adversary. But these problems are efficiently solvable on a large quantum computer using Shor's algorithm. Consequently, large-scale quantum computers would break the current authentication mechanisms of blockchains, leading to forgery of recorded transactions.

The goal of post-quantum cryptography (PQC) is to develop cryptographic algorithms that are safe from both classical and quantum attackers. The U.S. National Institute of Standards and Technology (NIST) chose CRYSTALS-Dilithium (now ML-DSA) and Falcon as the main lattice-based digital signature schemes after considerable work in recent years to standardize PQC \cite{Ducas_Kiltz_Lepoint_Lyubashevsky_Schwabe_Seiler_Stehlé_2018,cryptoeprint:2024/1769}. At the same time, there is growing interest in understanding how these schemes can be deployed in blockchain protocols \cite{article,cryptoeprint:2022/1151,10288193}. The ideas range from drop-in replacements for existing signatures in Bitcoin \cite{inbook} to constructions tailored specificaly for blockchain designs \cite{cryptoeprint:2022/1151}.

Majority of the literature on PQ signatures focuses on cryptographic properties and micro-benchmarks, for example, measuring key generation, signing, and verification costs in standalone libraries \cite{Ducas_Kiltz_Lepoint_Lyubashevsky_Schwabe_Seiler_Stehlé_2018,cryptoeprint:2024/1769}. These results are essential but they do not directly answer how different schemes affect the behaviour of an actual blockchain implementation. In a real system, the cost of signature depends on transaction, block size, and validation latency, and schemes with similar micro-benchmark performance may have different results when deployed at scale \cite{article,10288193}.

In this report, we study quantum-secure signatures within a blockchain environment. We implement a local, single-node blockchain prototype in which the underlying digital signature algorithm can be swapped between three lattice-based schemes: Dilithium, Falcon, and Hawk \cite{Ducas_Kiltz_Lepoint_Lyubashevsky_Schwabe_Seiler_Stehlé_2018,cryptoeprint:2024/1769,huanghttps}. The prototype implements a single wallet and transaction model, ensuring that modifications to the scheme do not impact the application logic. This provides a fair comparison of the algorithms from a blockchain perspective, measuring how they influence transaction signing cost, block verification time, and storage overhead via key and signature sizes.

The contributions of this preliminary report are as follows:
\begin{itemize}
    \item We illustrate a blockchain-like prototype that enables transparent switching between quantum-secure algorithms by separating application logic from the underlying digital signature technique.
    \item We outline a measuring approach for gathering micro-benchmark data (key generation, signature, verification) and combining it with blockchain-level metrics like validation latency and block creation.
    \item We place our work in the context of current research on digital signatures and post-quantum blockchains, and we identify how our prototype can be extended to additional schemes such as HAETAE \cite{cryptoeprint:2023/624}.
\end{itemize}

The full source code of the prototype blockchain and benchmarking scripts are available online.\footnote{\url{https://github.com/Stanotron/Post-Quantum-Blockchain}}
The outline of this report is as follows, Section~II reviews blockchain fundamentals. Section III introduces the post-quantum signature algorithms used in our prototype. Section IV describes the system design and measurement methodology, and Section V presents preliminary performance results. Finally, we conclude with a discussion and future work.

\section{Background}
This section briefly introduces blockchain fundamentals and the role of digital signatures, outlines the quantum threat to current schemes, and presents an overview of the post-quantum signatures considered in our report. We focus on concepts needed to understand the design of our blockchain prototype and the performance trade-offs between different algorithms.

\subsection{Blockchain and Digital Signatures}
A blockchain can be viewed as a replicated, append-only list maintained by a set of nodes that do not fully trust each other \cite{Nofer2017,8024092}. The list is organised into \emph{blocks}, each containing a batch of \emph{transactions}. Blocks are chained together using cryptographic hashes, so that modifying a past block would require recomputing all subsequent hashes, making tampering easily detectable. A protocol (such as proof-of-work or proof-of-stake) determines which blocks are accepted into the chain \cite{Nofer2017,buterin2013ethereum}.

Digital signatures play a key role in blockchain security. Each user controls one or more public-private key pairs, the public key (or its hash) acts as an address, and the private key is used to sign transactions that transfer value or currency \cite{inbook,buterin2013ethereum}. Nodes verify signatures before accepting transactions into their mempool or including them in a block. Similarly, some designs sign block headers such that a block can only be proposed by a validator or authorized miner. If an adversary can forge signatures, they can impersonate users, create fraudulent transactions, or rewrite parts of the chain.

In classical blockchains such as Bitcoin and Ethereum, signatures are typically instantiated with elliptic-curve schemes (e.g., ECDSA or Schnorr over secp256k1) \cite{inbook,buterin2013ethereum}. These schemes are efficient and have small key and signature sizes, which is attractive for systems where bandwidth and storage are constrained. However, their security relies on problems such as the elliptic-curve discrete logarithm problem, which are vulnerable to quantum attacks. This creates a long-term risk for any blockchain whose security guarantees must remain valid for decades, particularly given that transaction data is publicly stored and may be harvested today and attacked once quantum computers are available \cite{Nofer2017,10288193}.

\subsection{Quantum Threat and Post-Quantum Signatures}
Quantum computing poses a threat to widely deployed public-key cryptography. Shor's algorithm shows that integer factorisation and discrete logarithms can be solved in polynomial time on a large-scale quantum computer, breaking RSA and elliptic-curve schemes that are used in most current blockchain deployments \cite{10288193,cryptoeprint:2022/1151}. Even though practical, large-scale quantum computers do not yet exist, the long-term nature of blockchain data means that adversaries can perform ``store now, decrypt or forge later'' attacks, storing today's transaction history and exploiting it once quantum machines become available.

To address this threat, the cryptographic community has developed \emph{post-quantum} or \emph{quantum-resistant} schemes that are believed to be secure against both classical and quantum adversaries. The NIST Post-Quantum Cryptography standardization project has selected several algorithms for standardization, including lattice-based key encapsulation mechanisms and digital signatures \cite{10288193}. For signatures, NIST chose CRYSTALS-Dilithium (standardised as ML-DSA) and Falcon as its primary lattice-based candidates \cite{Ducas_Kiltz_Lepoint_Lyubashevsky_Schwabe_Seiler_Stehlé_2018,cryptoeprint:2024/1769}. These techniques are based on mathematical problems such as Module-LWE and NTRU, which currently have no known efficient quantum attacks and have undergone intense cryptanalytic inspection. Blockchains are a natural target application for PQ signatures, and a number of papers have been written studying how to migrate existing systems or design PQ-native protocols \cite{inbook,article,10288193,cryptoeprint:2022/1151}. 

In this report, we focus on four lattice-based digital signature schemes: CRYSTALS-Dilithium (ML-DSA), Falcon, Hawk, and HAETAE. All four are designed to be secure against quantum adversaries while offering acceptable performance on conventional hardware, but they exhibit different trade-offs in terms of key sizes, signature sizes, and computational cost. 

CRYSTALS-Dilithium is a lattice-based signature scheme built on the Module-LWE and Module-LWR problems \cite{Ducas_Kiltz_Lepoint_Lyubashevsky_Schwabe_Seiler_Stehlé_2018}. Falcon is another lattice-based signature scheme, based on the NTRU problem and using a more complex structure inspired by GPV signatures and fast Fourier sampling \cite{cryptoeprint:2024/1769}. Hawk is a more recent lattice-based signature scheme designed to provide short signatures and competitive performance \cite{huanghttps}. It targets similar application domains as Dilithium and Falcon, but with different design choices aimed at improving efficiency and signature size. HAETAE is another lattice-based proposal that aims to provide shorter Fiat--Shamir signatures with strong security guarantees \cite{cryptoeprint:2023/624}. 

In summary, ML-DSA, Falcon, Hawk, and HAETAE represent different points in the design space of lattice-based signatures. The main objective of the experiment outlined in the following sections is to understand how these trade-offs appear in a blockchain environment.

\subsection{Number Theoretic Transform in Lattice-Based Signatures}
\label{subsec:ntt-background}

Lattice-based schemes such as CRYSTALS-Dilithium, Falcon, Hawk, and HAETAE all work with polynomials over finite rings, for example $\mathbb{Z}_q[x]/(x^n+1)$, where $q$ is a prime modulus and $n$ is the polynomial degree \cite{Ducas_Kiltz_Lepoint_Lyubashevsky_Schwabe_Seiler_Stehlé_2018,cryptoeprint:2024/1769,huanghttps,cryptoeprint:2023/624}.  
In these schemes, the dominant operation is multiplying polynomials of length $n$, basically combining the coefficient vectors in a way so that each output coefficient depends on many input coefficients. A naive multiplication needs $O(n^2)$ modular multiplications and additions, which quickly becomes a problem for large $n$.

The Number Theoretic Transform (NTT) is the Discrete Fourier Transform defined over the ring $\mathbb{Z}_q$ instead of the complex numbers. By evaluating a polynomial at special points called roots of unity modulo $q$, multiplying the resulting values pointwise, and then applying the inverse transform, one can complete these operations in quasi-linear time $O(n \log n)$ instead of $O(n^2)$ \cite{cryptoeprint:2024/585}.

\section{Post-Quantum Signature Algorithms}
In this section we describe the four lattice-based signature schemes considered in this work: CRYSTALS-Dilithium (standardised as ML-DSA), Falcon, Hawk, and HAETAE.

\subsection{CRYSTALS-Dilithium / ML-DSA}

CRYSTALS-Dilithium is a lattice-based digital signature scheme whose security is based on the hardness of finding short vectors in module lattices, expressed using the Module-LWE and Module-SIS problems \cite{Ducas_Kiltz_Lepoint_Lyubashevsky_Schwabe_Seiler_Stehlé_2018}. The scheme was later standardised by NIST under the name ML-DSA, and is considered as a long-term replacement for current signatures such as RSA and ECDSA in a post-quantum setting.

Dilithium operates over the polynomial ring
\[
R_q = \mathbb{Z}_q[X]/(X^n + 1),
\]
where $n = 256$ and $q = 8380417$ is a prime chosen so that fast NTT-based polynomial multiplication is possible \cite{Ducas_Kiltz_Lepoint_Lyubashevsky_Schwabe_Seiler_Stehlé_2018}. Computations are expressed as linear algebra over $R_q$ (matrices and vectors with polynomial entries), with polynomial products accelerated using the NTT.

\paragraph{High-level algorithms.}
For presentation and later analysis, we summarise the scheme as a compact \textsf{KeyGen}/\textsf{Sign}/\textsf{Verify} view. (The real specification additionally includes compression, ``high/low bits'' decomposition, hint values, and rejection sampling; these details do not change the overall algebraic structure shown below \cite{Ducas_Kiltz_Lepoint_Lyubashevsky_Schwabe_Seiler_Stehlé_2018}.)

\subsubsection*{Key generation (high level)}
\begin{enumerate}
  \item Sample a public matrix $A \in_R R_q^{k\times \ell}$ (from a short seed).
  \item Sample short secrets $s_1 \in_R \mathcal{S}_\eta^{\ell}$ and $s_2 \in_R \mathcal{S}_\eta^{k}$.
  \item Compute
  \[
    t = A s_1 + s_2 \in R_q^k \pmod q.
  \]
  \item Output public and secret keys:
  \[
    \mathrm{pk}=(A,t), \qquad \mathrm{sk}=(s_1,s_2)\,,
  \]
  where in the instantiated scheme $\mathrm{pk}$ and $\mathrm{sk}$ also store auxiliary hashes and compressed encodings \cite{Ducas_Kiltz_Lepoint_Lyubashevsky_Schwabe_Seiler_Stehlé_2018}.
\end{enumerate}
Recovering $s_1$ from $(A,t)$ corresponds to a Module-LWE instance \cite{Ducas_Kiltz_Lepoint_Lyubashevsky_Schwabe_Seiler_Stehlé_2018}.

\subsubsection*{Signature generation (high level)}
To sign a message $M \in \{0,1\}^\ast$:
\begin{enumerate}
  \item Sample a random masking vector $y \in_R \mathcal{S}_{\gamma_1}^{\ell}$.
  \item Compute a commitment $w = A y$ (in the real scheme, the verifier-relevant part is derived from the high bits of $w$) \cite{Ducas_Kiltz_Lepoint_Lyubashevsky_Schwabe_Seiler_Stehlé_2018}.
  \item Compute the challenge $c = H(M \,\|\, w)$, where $c$ has bounded weight/structure as specified \cite{Ducas_Kiltz_Lepoint_Lyubashevsky_Schwabe_Seiler_Stehlé_2018}.
  \item Compute the response
  \[
    z = y + c s_1.
  \]
  \item Output the signature $\sigma=(c,z)$, together with the additional hint/compression information required by the concrete ML-DSA encoding \cite{Ducas_Kiltz_Lepoint_Lyubashevsky_Schwabe_Seiler_Stehlé_2018}.
\end{enumerate}

\subsubsection*{Signature verification (high level)}
To verify $\sigma=(c,z)$ on $M$ using the authentic public key $(A,t)$:
\begin{enumerate}
  \item Recompute
  \[
    w' = A z - c t \pmod q.
  \]
  \item Accept iff the recomputed challenge matches:
  \[
    c = H(M \,\|\, w'),
  \]
  and the scheme-specific bound checks and hint-assisted reconstruction succeed \cite{Ducas_Kiltz_Lepoint_Lyubashevsky_Schwabe_Seiler_Stehlé_2018}.
\end{enumerate}

In our blockchain prototype we use the three ML-DSA parameter sets exposed by the \texttt{liboqs} \cite{liboqs} library, corresponding to ML-DSA-44, ML-DSA-65, and ML-DSA-87. All three share the same ring dimension $n=256$ and modulus $q=8380417$, but differ in matrix dimensions $(k,\ell)$ and bounds such as $\eta$, $\gamma_1$, $\gamma_2$, $\beta$, and $\omega$, which together determine security level and efficiency \cite{Ducas_Kiltz_Lepoint_Lyubashevsky_Schwabe_Seiler_Stehlé_2018}. Higher security levels use larger matrices, which directly increases the number of polynomial multiplications and the size of public keys and signatures.

Table~\ref{tab:mldsa-params} summarises the main parameters we rely on in the rest of the report.

\begin{table}[t]
\caption{ML-DSA parameter sets used in our experiments \cite{Ducas_Kiltz_Lepoint_Lyubashevsky_Schwabe_Seiler_Stehlé_2018}.}
\label{tab:mldsa-params}
\centering
\begin{tabular}{lrrr}
\hline
Parameter & ML-DSA-44 & ML-DSA-65 & ML-DSA-87 \\
\hline
$(k,\ell)$                 & (4, 3)  & (6, 4)  & (8, 5)  \\
PK size [bytes]            & 1312    & 1952    & 2592    \\
SK size [bytes]            & 2560    & 4032    & 4896    \\
Signature size [bytes]     & 2420    & 3309    & 4627    \\
\hline
\end{tabular}
\end{table}

\subsection{Falcon}

Falcon is a lattice-based digital signature scheme built on the NTRU problem and the GPV hash-and-sign framework \cite{cryptoeprint:2024/1769}. Conceptually, the public key defines a linear relation over a structured lattice, while the secret key provides a trapdoor that enables efficient sampling of a short preimage for hashed message targets. Verification checks both correctness of the relation and that the signature is short.

Falcon works over the NTRU-style ring
\[
  R_q = \mathbb{Z}_q[X]/(X^n + 1),
\]
where $n$ is a power of two and $q$ is a small prime modulus \cite{cryptoeprint:2024/1769}. The NIST parameter sets use $n \in \{512, 1024\}$ and modulus $q = 12289$ \cite{cryptoeprint:2024/1769}, enabling fast polynomial arithmetic using FFT/NTT-style techniques. In practice, Falcon relies on a specialised discrete Gaussian sampler (via Fast Fourier Orthogonalization) to produce short signatures with the right distribution \cite{cryptoeprint:2024/1769}.

\paragraph{High-level algorithms.}
To align with our implementation-level discussion later, we summarise Falcon using a compact \textsf{KeyGen}/\textsf{Sign}/\textsf{Verify} view. (The real scheme includes encoding/compression and restart conditions; the structure below captures the core relations \cite{cryptoeprint:2024/1769}.)

\subsubsection*{Key generation (high level)}
\begin{enumerate}
  \item Sample short polynomials $f,g,F,G \in \mathbb{Z}[X]/(X^n+1)$ such that the NTRU equation holds:
  \[
    fG - gF = q.
  \]
  \item Construct a short basis (trapdoor) for the corresponding NTRU lattice:
  \[
    B =
    \begin{pmatrix}
      g & -f \\
      G & -F
    \end{pmatrix}.
  \]
  \item Compute the public key polynomial:
  \[
    h = g f^{-1} \bmod q.
  \]
  \item Output:
  \[
    \mathrm{pk}=h, \qquad \mathrm{sk}=B,
  \]
  where $\mathrm{pk}$ may also include an encoding of $h$ and $\mathrm{sk}$ includes auxiliary data for sampling \cite{cryptoeprint:2024/1769}.
\end{enumerate}

\subsubsection*{Signature generation (high level)}
To sign a message $m$:
\begin{enumerate}
  \item Hash the message to a target in the public relation:
  \[
    H(m).
  \]
  \item Using the trapdoor basis $B$, sample a lattice vector $v \in \Lambda(B)$ that is close to the target representation.
  \item Compute the short preimage (the signature vector):
  \[
    s = c - v = (s_1,s_2),
  \]
  where $c$ is the target derived from $H(m)$ in the public-key relation.
  \item Output the signature $\sigma = s$, together with the scheme’s compressed encoding (Falcon signatures are variable-length due to compression and possible restarts) \cite{cryptoeprint:2024/1769}.
\end{enumerate}

\subsubsection*{Signature verification (high level)}
Given public key $h$, message $m$, and signature $s=(s_1,s_2)$:
\begin{enumerate}
  \item Check that $s$ is \textbf{short} (norm bound).
  \item Check the public relation (NTRU-lattice equation):
  \[
    sA = H(m) \pmod q,
  \]
  for the public map $A$ derived from $h$ (commonly written as $A = (1, h)^T$ in compact matrix form).
  \item Accept iff both checks hold \cite{cryptoeprint:2024/1769}.
\end{enumerate}

The official Falcon parameter sets, Falcon-512 and Falcon-1024, target NIST security levels I and V, respectively. Both share the same modulus but differ in ring dimension and internal parameters that control Gaussian sampling and the maximum allowed signature norm. Table~\ref{tab:falcon-params} summarises the main public parameters taken from the Falcon security analysis \cite{cryptoeprint:2024/1769}.

\begin{table}[t]
\caption{Selected Falcon parameter values (from \cite{cryptoeprint:2024/1769}).}
\label{tab:falcon-params}
\centering
\begin{tabular}{lrr}
\hline
Parameter & Falcon-512 & Falcon-1024 \\
\hline
NIST security level          & I       & V \\
Ring degree $n$              & 512     & 1024 \\
Modulus $q$                  & 12289   & 12289 \\
Std.\ deviation $s$ (approx.)& 165.74  & 168.39 \\
Max.\ signature norm bound $\beta$ & 5833.93 & 8382.44 \\
Salt length $k$ (bits)       & 320     & 320 \\
\hline
\end{tabular}
\end{table}

\noindent
In our prototype and measurements, Falcon’s main practical advantage is its compact public keys and signatures, which reduces transaction and block size. The main trade-off is implementation complexity and noticeably higher key generation cost compared to ML-DSA in our test environment \cite{cryptoeprint:2024/1769}.

\subsection{HAWK}

Hawk is a lattice-based digital signature scheme whose security is based on problems from the lattice isomorphism problem (LIP) framework \cite{huanghttps,cryptoeprint:2022/1151}. It follows a hash-and-sign approach: a message is hashed into a structured target, and the signer uses a trapdoor to produce a short lattice vector that satisfies a public relation. The current Hawk specification (version~1.1) defines several parameter sets, of which HAWK-512 and HAWK-1024 are intended to provide NIST level I and level V security, respectively \cite{huanghttps}.

At a high level, Hawk operates over a polynomial ring of the form
\[
R_q = \mathbb{Z}_q[X]/(X^n + 1),
\]
where $n$ is a power of two and $q$ is a prime modulus chosen to support efficient polynomial arithmetic \cite{huanghttps}. The parameter sets HAWK-512 and HAWK-1024 use degrees $n \in \{512,1024\}$; this supports fast structured polynomial arithmetic (e.g., NTT-style multiplication) in implementations \cite{huanghttps}.

\paragraph{High-level algorithms.}
To align with our later engineering and benchmarking discussion, we summarise Hawk using a compact \textsf{KeyGen}/\textsf{Sign}/\textsf{Verify} view. (The full specification includes additional details such as symmetry breaking, rejection conditions, and exact encoding/compression; these do not change the overall flow below \cite{huanghttps}.)

\subsubsection*{Key generation (high level)}
\begin{enumerate}
  \item Sample $f,g \in R_n$ with i.i.d.\ coefficients:
  \[
    f[i],g[i] \leftarrow \mathrm{Bin}(\eta).
  \]
  \item Find $(F,G)\in R_n^2$ such that the NTRU relation holds:
  \[
    fG - gF = 1 \quad \text{in } R_n.
  \]
  \item Define the secret trapdoor basis:
  \[
    B=\begin{pmatrix} f & F \\ g & G \end{pmatrix}.
  \]
  \item Define the public key matrix (public quadratic form):
  \[
    Q = B^{\star}B.
  \]
  \item Hash the public key:
  \[
    h_{\mathrm{pub}} = H(Q).
  \]
  \item Output:
  \[
    \mathrm{pk}=Q,\qquad \mathrm{sk}=(B, h_{\mathrm{pub}}).
  \]
\end{enumerate}

\subsubsection*{Signature generation (high level)}
To sign a message $m$ with $\mathrm{sk}=(B,h_{\mathrm{pub}})$:
\begin{enumerate}
  \item Compute \(M \leftarrow H(m \,\|\, h_{\mathrm{pub}})\).
  \item Sample \(\mathrm{salt}\leftarrow\{0,1\}^{\lambda}\).
  \item Compute \(h \leftarrow H(M \,\|\, \mathrm{salt}) \in R_n^2\).
  \item Sample a short vector $x$ from a Gaussian-like distribution; reject and retry if $\|x\|$ is too large.
  \item Compute:
  \[
    w \leftarrow B^{-1}x,
    \qquad
    s \leftarrow \tfrac{1}{2}(h-w).
  \]
  \item Output the signature:
  \[
    \mathrm{sig}=(\mathrm{salt},\, s_1),
  \]
  where $s_1$ is a compressed encoding derived from $s$ \cite{huanghttps}.
\end{enumerate}

\subsubsection*{Signature verification (high level)}
Given $\mathrm{pk}=Q$, message $m$, and $\mathrm{sig}=(\mathrm{salt},s_1)$:
\begin{enumerate}
  \item Recompute \(h_{\mathrm{pub}} \leftarrow H(Q)\).
  \item Compute \(M \leftarrow H(m \,\|\, h_{\mathrm{pub}})\).
  \item Compute \(h \leftarrow H(M \,\|\, \mathrm{salt}) \in R_n^2\).
  \item Reconstruct $s=(s_0,s_1)$ from $(Q,h,s_1)$ and compute:
  \[
    w \leftarrow h - 2s.
  \]
  \item Accept if $w$ is valid (including symmetry-breaking consistency) and satisfies the norm bound:
  \[
    \|w\|_{Q} \le \text{bound}.
  \]
\end{enumerate}

Like the other lattice-based schemes considered in this work, Hawk relies heavily on structured polynomial arithmetic and fast hashing. The reference implementation is written in constant-time C, uses only integer arithmetic, and supports both a generic and an AVX2-optimised version \cite{huanghttps}. Table~\ref{tab:hawk-params} summarises the main public parameter sets relevant for our experiments. In addition to HAWK-512 and HAWK-1024, the specification also defines a lower-security ``challenge'' set HAWK-256 intended mainly for cryptanalysis rather than deployment \cite{huanghttps}. In this work we focus on HAWK-512 and HAWK-1024.

\begin{table}[t]
\caption{Hawk parameter sets (from \cite{huanghttps}).}
\label{tab:hawk-params}
\centering
\begin{tabular}{lrr}
\hline
Parameter               & HAWK-512 & HAWK-1024 \\
\hline
Degree $n$              & 512      & 1024      \\
PK size [bytes]         & 1024     & 2440      \\
SK size [bytes]         & 184      & 360       \\
Signature size [bytes]  & 555      & 1221      \\
\hline
\end{tabular}
\end{table}


\noindent
In our blockchain prototype, Hawk is used as an additional PQ signature option alongside ML-DSA and Falcon, allowing us to compare the effects of different lattice-based design choices on signature sizes and per-transaction verification costs.

\subsection{HAETAE}

HAETAE is a lattice-based digital signature scheme that follows the Fiat--Shamir-with-aborts paradigm, like Dilithium, but is designed to reduce signature and verification key sizes for space-constrained applications such as IoT devices \cite{cryptoeprint:2023/624}. Its security is based on module variants of the Learning With Errors (MLWE) and Short Integer Solution (MSIS) problems in the Quantum Random Oracle Model, and the parameter sets are chosen using a Core-SVP style security analysis \cite{cryptoeprint:2023/624}.

HAETAE works over the polynomial ring
\[
R = \mathbb{Z}_{q}[X]/(X^n + 1),
\]
with degree $n = 256$ and modulus $q = 64513$ for all parameter sets \cite{cryptoeprint:2023/624}. These choices are NTT-friendly, so polynomial multiplications can be implemented efficiently using NTT and inverse NTT. The public key contains a structured public matrix $A$, while the secret key contains a short vector $s$; the module dimensions $(k,\ell)$ increase with the targeted security level.

\paragraph{High-level algorithms.}
To match the level of detail used for ML-DSA and Falcon, we summarise HAETAE using a compact \textsf{KeyGen}/\textsf{Sign}/\textsf{Verify} view. (The full specification includes additional steps such as compression/entropy coding, auxiliary values, and multiple rejection checks; the structure below captures the main flow \cite{cryptoeprint:2023/624}.)

\subsubsection*{Key generation (high level)}
\begin{enumerate}
  \item Sample $A_{\text{gen}} \leftarrow R_q^{k\times(\ell-1)}$ and small secrets $s_{\text{gen}}, e_{\text{gen}}$.
  \item Compute
  \[
    b \leftarrow A_{\text{gen}} s_{\text{gen}} + e_{\text{gen}} \in R_q^k.
  \]
  \item Build the public matrix $A$ from $(A_{\text{gen}}, b)$.
  \item Form the secret vector
  \[
    s \leftarrow (1,\, s_{\text{gen}}^\top,\, e_{\text{gen}}^\top)^\top.
  \]
  \item Restart if the secret fails the scheme’s bound checks (used to keep later signing efficient) \cite{cryptoeprint:2023/624}.
  \item Output:
  \[
    \mathrm{sk}=s,\qquad \mathrm{pk}=A.
  \]
\end{enumerate}

\subsubsection*{Signature generation (high level)}
To sign a message $M$:
\begin{enumerate}
  \item Sample a masking vector using the hyperball sampler:
  \[
    y \leftarrow \mathcal{U}(\text{Hyperball}).
  \]
  \item Compute the challenge
  \[
    c \leftarrow H(\textsf{HighBits}(A y,\alpha),\, \textsf{aux}(y),\, M),
  \]
  where $c$ is sparse with fixed weight (controlled by $\tau$) \cite{cryptoeprint:2023/624}.
  \item Choose $b\leftarrow\{0,1\}$ and compute the response
  \[
    z \leftarrow y + (-1)^b\, c\cdot s.
  \]
  \item Compute a helper/hint value $h$ (for high-bits consistency during verification):
  \[
    h \leftarrow \textsf{HighBits}(A z,\alpha) - \textsf{HighBits}(A_1 z_1,\alpha).
  \]
  \item Reject and retry if $\|z\|_2$ violates the bound (and apply one additional light rejection step) \cite{cryptoeprint:2023/624}.
  \item Output (packed/encoded in the concrete scheme):
\[
\begin{aligned}
\sigma = (&\textsf{Encode}(\textsf{HighBits}(z_1,\alpha)),\\
          &\textsf{LowBits}(z_1,\alpha),\ h,\ c).
\end{aligned}
\]
\end{enumerate}

\subsubsection*{Signature verification (high level)}
Given $\mathrm{pk}=A$, message $M$, and signature $\sigma$:
\begin{enumerate}
  \item Decode/reconstruct $\tilde{z}$ from $\sigma$ (using $h$ and the decomposition).
  \item Recompute auxiliary input $h' \leftarrow \textsf{aux}(\tilde{z})$.
  \item Recompute the challenge
  \[
    c' \leftarrow H(\textsf{HighBits}(A\tilde{z}-q c j,\alpha),\, h',\, M).
  \]
  \item Accept iff $c'=c$ and $\|\tilde{z}\|_2$ satisfies the verification bound \cite{cryptoeprint:2023/624}.
\end{enumerate}

HAETAE comes with three main parameter sets, denoted HAETAE-120, HAETAE-180, and HAETAE-260, which roughly target NIST security levels 2, 3, and 5. All three share the same ring degree $n = 256$ and modulus $q = 64513$, but differ in $(k,\ell)$, the challenge weight $\tau$, and the rejection bounds. Table~\ref{tab:haetae-params} summarises the main parameters we rely on in the rest of the report.

\begin{table}[t]
\caption{HAETAE parameter sets (from \cite{cryptoeprint:2023/624}).}
\label{tab:haetae-params}
\centering
\begin{tabular}{lrrr}
\hline
Parameter & HAETAE-120 & HAETAE-180 & HAETAE-260 \\
\hline
$(k,\ell)$                 & (2, 4)    & (3, 6)    & (4, 7)    \\
PK size [bytes]            & 992       & 1472      & 2080      \\
SK size [bytes]            & 1408      & 2112      & 2752      \\
Signature size [bytes]     & 1474      & 2349      & 2948      \\
\hline
\end{tabular}
\end{table}

\noindent
In this work, we treat HAETAE as an additional post-quantum signature candidate for comparison. We use the authors' reference implementation to obtain standalone microbenchmark results for key generation, signing, and verification, but we do not integrate HAETAE into our Windows-based blockchain prototype. The reference code relies on GCC-specific features such as variable-length arrays and uses OpenSSL in its random number generator, which does not compile cleanly under our MSVC toolchain without modifying the code or adding extra dependencies. As a result, HAETAE is evaluated only through standalone microbenchmarks rather than full blockchain integration, but these results still allow us to compare its performance and size characteristics with the other schemes in Section~IV.

\section{System Design and Methodology}

This section describes the software architecture of our prototype blockchain, how post-quantum signature schemes are integrated, and how we obtain the performance measurements reported in Section~V. Our goal is to isolate the impact of the digital signature algorithm on transaction and block processing while keeping the rest of the system as simple and uniform as possible.

\subsection{Experimental Environment}

All experiments were run on an x86\_64 desktop machine running a 64-bit Windows operating system. The prototype was compiled in Release mode using the Microsoft Visual C++ toolchain, and linked against the \texttt{liboqs} \cite{liboqs} library for post-quantum signatures. We used the \texttt{ML\_DSA\_44}, \texttt{ML\_DSA\_65}, and \texttt{ML\_DSA\_87} parameter sets for CRYSTALS-Dilithium (standardised as ML-DSA), and the \texttt{Falcon-512} and \texttt{Falcon-1024} parameter sets for Falcon, in line with the NIST post-quantum cryptography process \cite{Ducas_Kiltz_Lepoint_Lyubashevsky_Schwabe_Seiler_Stehlé_2018,cryptoeprint:2024/1769,10288193}. For Hawk, we used the authors' reference implementation since hawk is not a part of \texttt{liboqs} library\cite{huanghttps, hawk-dev}.

Timing measurements are taken using a high-resolution wall-clock timer returning microsecond timestamps. To reduce noise, we execute each micro-benchmark (e.g., key generation or block validation) a fixed number of times and report the average over all iterations. 
In particular, block validation time is obtained by repeatedly calling \texttt{validateBlock} on a fully populated block, rather than extrapolating from microbenchmark results.
All results shown in Section~V are based on 100 key generations, 1000 transaction signatures within a block, and 100 repeated block validations for each algorithm.

\subsection{Blockchain Prototype}

The prototype follows an account-based model similar to Ethereum \cite{buterin2013ethereum}, but omits consensus and networking to focus on the cryptographic overhead. The core data structures are:

\begin{itemize}
    \item \textbf{Transaction:} A transaction contains a sender public key, a recipient public key, a transfer amount, a monotonically increasing nonce, and a digital signature over the transaction body (all fields except the signature itself).
    \item \textbf{Block:} A block consists of an index, the hash of the previous block, a timestamp, a vector of transactions, and the hash of the block contents. The block hash is computed over the serialized header and all transactions (including their signatures).
    \item \textbf{Blockchain:} The blockchain maintains a vector of blocks, starting with a hard-coded block. The prototype provides methods to create a new block from a batch of transactions and to validate a block.
    \item \textbf{Wallet:} A wallet encapsulates a public--private key pair for a given signature scheme and exposes a method to create and sign transactions to a recipient public key.
\end{itemize}

For hashing, we use SHA3-256 hash function used only to define block identifiers and block linkages. This function is intended to be cryptographically secure and is sufficient for our purpose of measuring serialization overhead and verifying that tampering with block contents is detected by recomputing the block hash.

The \texttt{Blockchain::validateBlock} method checks: (i) that the block index and \texttt{prev\_hash} correctly link to the latest block in the chain; (ii) that the block hash matches a recomputation over the serialized contents; and (iii) that every transaction signature in the block verifies under the corresponding sender public key. If any check fails, the block is rejected.

The full source code of the prototype blockchain and the benchmarking scripts used in this work are available in an open-source repository.\footnote{\url{https://github.com/Stanotron/Post-Quantum-Blockchain}}

\subsection{Cryptographic Integration}

To abstract over concrete signature schemes, we define a \texttt{Crypto} interface with the following core operations:

\begin{itemize}
    \item \texttt{generateKeypair():} produce a public/secret key pair;
    \item \texttt{sign(msg, sk):} sign a message using a secret key;
    \item \texttt{verify(msg, sig, pk):} verify a signature against a message and public key;
    \item metadata methods such as \texttt{name()}, \texttt{family()}, \texttt{variant()}, and functions returning expected public key, secret key, and maximum signature sizes.
\end{itemize}

Concrete implementations of this interface wrap different backends:

\begin{itemize}
    \item \textbf{OqsMldsaCrypto:} wraps the \texttt{liboqs} APIs for ML-DSA (Dilithium) and supports variants 44, 65, and 87 \cite{liboqs}.
    \item \textbf{OqsFalconCrypto:} wraps the \texttt{liboqs} APIs for Falcon, supporting Falcon-512 and Falcon-1024 \cite{liboqs}.
    \item \textbf{HawkCrypto:} wraps the reference implementation of Hawk and presents it through the same interface \cite{hawk-dev}.
\end{itemize}

An \texttt{AlgoConfig} structure encodes the selected algorithm family (\texttt{ML\_DSA}, \texttt{FALCON}, or \texttt{HAWK}) and variant (e.g., \texttt{"44"}, \texttt{"512"}, or \texttt{"1024"}). A factory function \texttt{createCrypto} takes this configuration and instantiates the appropriate concrete \texttt{Crypto} implementation. The rest of the system wallets, transactions, and blocks only depend on the abstract \texttt{Crypto} interface, so swapping the signature scheme does not require changes to the application logic.

\subsection{Benchmark Workflow and Metrics}

The main benchmarking program performs the following steps for each selected algorithm:

\begin{enumerate}
    \item \textbf{Key generation benchmark:} Using the configured \texttt{Crypto} instance, we call \texttt{generateKeypair} a total of 100 times. Each call is timed individually, and we compute the average key generation time per wallet in microseconds.
    \item \textbf{Wallet setup:} We construct a \texttt{Blockchain} object bound to the same \texttt{Crypto} instance and create two wallets, ``Alice'' and ``Bob''. Each wallet generates a new key pair, and we record the size of their public keys.
    \item \textbf{Transaction signing benchmark:} We create 1000 transactions from Alice to Bob, with a fixed transfer amount and a monotonically increasing nonce. Each transaction is signed by Alice's wallet, which internally serializes the transaction body and invokes \texttt{Crypto::sign}. We measure the average signing time per transaction in this block context.
    \item \textbf{Block creation and size measurement:} The 1000 signed transactions are bundled into a single block on top of the genesis block. We compute the block hash, serialize the full block (including the hash), and record the total serialized block size and the average serialized size per transaction.
    \item \textbf{Block validation benchmark:} We first validate the block once to ensure it is accepted, then deliberately flip a bit in the first transaction's signature to confirm that validation fails as expected. Finally, we run 100 repeated validations of the original block, measuring the total time and deriving both the average time to validate the whole block and the average per-transaction verification time.
\end{enumerate}

From this workflow we obtain the following metrics for each scheme and parameter set:

\begin{itemize}
    \item average key generation time per wallet;
    \item average transaction signing time in a 1000-transaction block;
    \item serialized public key size, signature size, average transaction size, and block size;
    \item average block validation time and average per-transaction verification time.
\end{itemize}

These metrics allow us to compare schemes both at the \emph{micro} level (key generation, signing, verifying) and at the \emph{blockchain} level (impact on block size and validation latency), following the system-oriented perspective advocated in prior work on post-quantum blockchains \cite{article,10288193,cryptoeprint:2022/1151}.

\section{Experimental Results}

This section summarizes the measurements obtained from the benchmark described above. Each result corresponds to a block containing 1000 transactions from Alice to Bob, using the same transaction structure and serialization format across all algorithms. We report size metrics for keys, signatures, and blocks, followed by timing metrics for key generation, transaction signing, and transaction verification.

\subsection{Size Metrics}

Table~\ref{tab:sizes} reports the public key size, signature size, average serialized transaction size, and total serialized block size for a block with 1000 transactions for each scheme and parameter set. All sizes are given in bytes.

\begin{table}[t]
\caption{Size metrics for a 1000-transaction block (bytes).}
\label{tab:sizes}
\centering
\begin{tabular}{lrrrr}
\hline
Scheme & PK & Signature & Avg tx size & Block size \\
\hline
ML-DSA-44   & 1312 & 2420 & 5092 & 5092084 \\
ML-DSA-65   & 1952 & 3309 & 7261 & 7261084 \\
ML-DSA-87   & 2592 & 4627 & 9859 & 9859084 \\
Falcon-512  &  897 &  658 & 2497 & 2497118 \\
Falcon-1024 & 1793 & 1275 & 4905 & 4904685 \\
Hawk-512    & 1024 &  555 & 2651 & 2651084 \\
Hawk-1024   & 2440 & 1221 & 6149 & 6149084 \\
\hline
\end{tabular}
\end{table}

Within the ML-DSA family, increasing the security level from 44 to 87 leads to a monotonic increase in public key size, signature size, and transaction size. The average transaction size grows from approximately 5.1~kB to 9.9~kB, and the total block size (with 1000 transactions) grows from about 5.1~MB to 9.9~MB. This behaviour is consistent with the parameterisation of Dilithium, where higher security corresponds to larger underlying lattice dimensions and hence larger keys and signatures \cite{Ducas_Kiltz_Lepoint_Lyubashevsky_Schwabe_Seiler_Stehlé_2018}.

Falcon exhibits substantially smaller signatures than ML-DSA at comparable security levels: Falcon-512 signatures are only 658~bytes, compared to 2420~bytes for ML-DSA-44, and Falcon-1024 signatures are 1275~bytes, compared to 4627~bytes for ML-DSA-87. As a result, Falcon-based transactions and blocks are significantly more compact; for example, the Falcon-512 block with 1000 transactions is roughly 2.5~MB, versus over 5~MB for ML-DSA-44. This aligns with the design goals of Falcon, which targets compact public keys and signatures using NTRU lattices and carefully optimised sampling \cite{cryptoeprint:2024/1769}.

Hawk offers even shorter signatures than Falcon at both parameter sets, with 555~bytes for Hawk-512 and 1221~bytes for Hawk-1024. Public keys for Hawk are somewhat larger than Falcon's at the same security level, but still smaller than those of ML-DSA. Consequently, Hawk-based transactions and blocks are also compact: Hawk-512 yields an average transaction size of about 2.65~kB and a block size of about 2.65~MB. These results highlight Hawk as an attractive candidate from a bandwidth and storage perspective, though it is not yet standardised and still under active evaluation \cite{huanghttps,cryptoeprint:2022/1151}.

\begin{figure}[t]
\centering
\includegraphics[width=\linewidth]{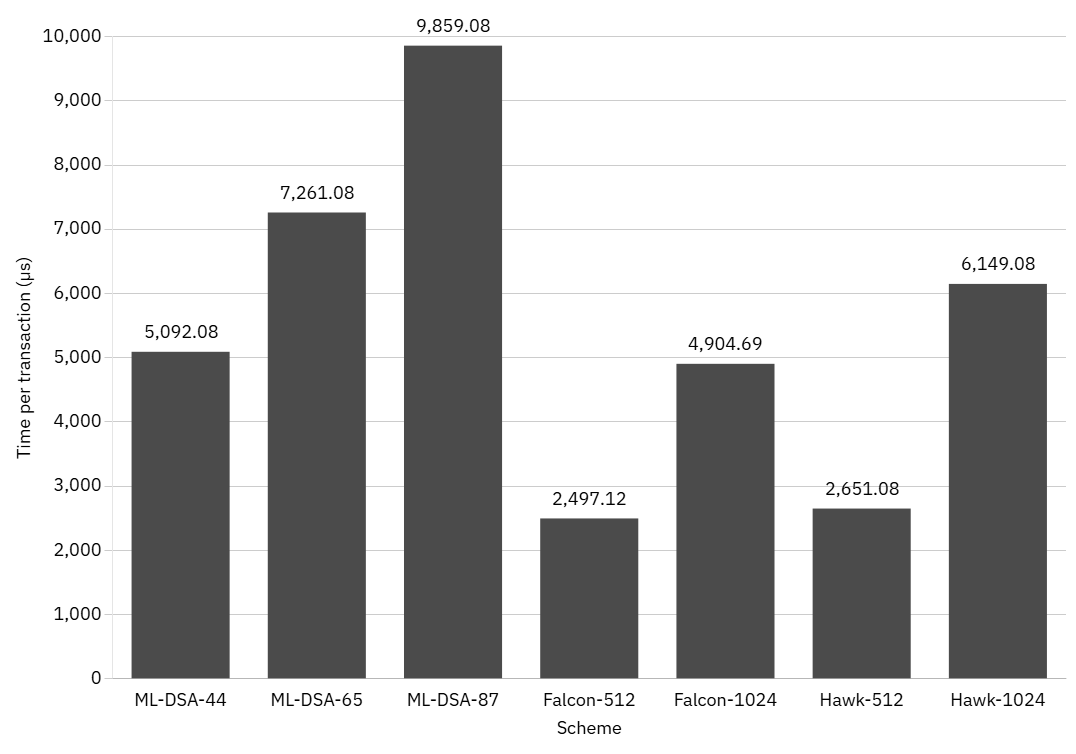}
\caption{Average serialized transaction size for each signature scheme and parameter set (1000-transaction block).}
\label{fig:tx-size}
\end{figure}

Figure~\ref{fig:tx-size} can be used to visualise the average transaction size per scheme as a bar chart, making the trade-offs between different parameter sets immediately apparent.

\subsection{Timing Metrics}

Table~\ref{tab:timings} presents timing results for each scheme: average key generation time per wallet, average transaction signing time in the block context, and average per-transaction verification time derived from repeated block validation. All values are in microseconds (µs).

\begin{table}[t]
\caption{Timing metrics (microseconds, average over all iterations).}
\label{tab:timings}
\centering
\begin{tabular}{lrrr}
\hline
Scheme & Keygen & Sign / tx & Verify / tx \\
\hline
ML-DSA-44   &    78.9 &   281.3 &   97.5 \\
ML-DSA-65   &   138.4 &   457.8 &  144.5 \\
ML-DSA-87   &   200.3 &   570.9 &  234.1 \\
Falcon-512  &  5721.4 &   206.1 &   51.6 \\
Falcon-1024 & 16825.6 &   401.9 &  103.5 \\
Hawk-512    &  1477.8 &    31.4 &   47.1 \\
Hawk-1024   &  9549.2 &    70.0 &   98.5 \\
\hline
\end{tabular}
\end{table}

Within the ML-DSA family, higher security levels again translate into higher computational cost. Average transaction signing time increases from approximately 0.28~ms (ML-DSA-44) to 0.57~ms (ML-DSA-87), while per-transaction verification time increases from about 0.10~ms to 0.23~ms. Key generation cost remains relatively low (well below 1~ms for all three variants), which is consistent with their suitability for frequent wallet creation in blockchain systems \cite{Ducas_Kiltz_Lepoint_Lyubashevsky_Schwabe_Seiler_Stehlé_2018}.

Falcon presents a somewhat different profile. Key generation is significantly more expensive---around 5.7~ms for Falcon-512 and 16.8~ms for Falcon-1024---reflecting the complexity of its lattice sampling procedures \cite{cryptoeprint:2024/1769}. However, signing and verification are comparatively fast: Falcon-512 signs a transaction in about 0.21~ms on average and verifies in about 0.052~ms, making it faster than all ML-DSA variants in terms of verification, which is the dominant operation for full validating nodes.

Hawk shows particularly strong performance for signing: Hawk-512 signs in roughly 0.03~ms and Hawk-1024 in about 0.07~ms per transaction, significantly faster than both ML-DSA and Falcon at similar security levels. Verification times are also competitive: Hawk-512 verifies in about 0.047~ms per transaction, slightly faster than Falcon-512, and Hawk-1024 verifies in about 0.10~ms, comparable to Falcon-1024. Key generation for Hawk lies between ML-DSA and Falcon, with Hawk-512 at roughly 1.5~ms and Hawk-1024 at around 9.6~ms.

\begin{figure}[t]
\centering
\includegraphics[width=\linewidth]{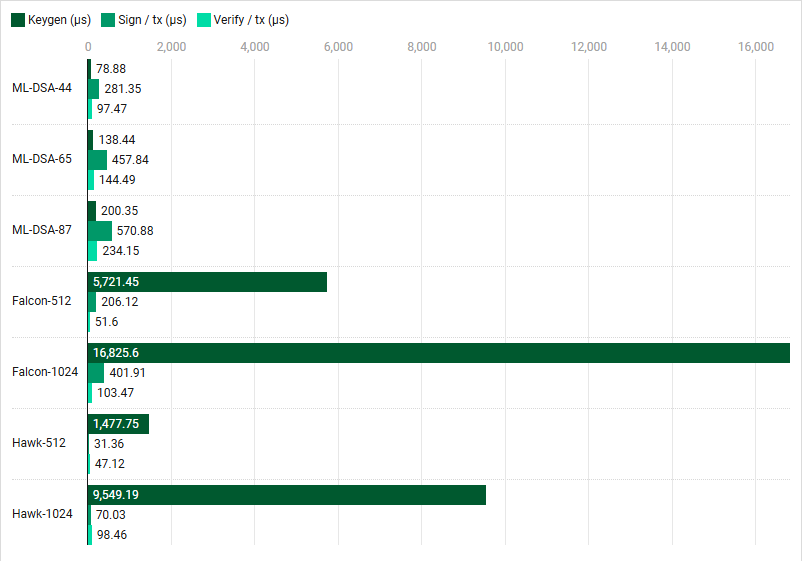}
\caption{Average key gen, signing and per-transaction verification times for each scheme (logarithmic or linear scale as preferred).}
\label{fig:sign-verify}
\end{figure}

Figure~\ref{fig:sign-verify} can be populated with a bar chart comparing signing and verification times across schemes, clearly illustrating the relative positions of ML-DSA, Falcon, and Hawk.

\subsection{Standalone HAETAE Microbenchmarks}

In addition to the integrated blockchain benchmarks, we also measured the standalone performance of HAETAE using the authors' reference implementation \texttt{haetae2-benchmark}, \texttt{haetae3-benchmark}, and \texttt{haetae5-benchmark} on the same machine and with the same high-resolution timing setup as described in Section~III. These measurements report the time for a single key generation, signature, and verification operation, averaged over many iterations, but do not include any blockchain-specific overhead such as transaction serialization or block validation. We therefore treat them as pure cryptographic microbenchmarks that complement the integrated results for ML-DSA, Falcon, and Hawk.

Table~\ref{tab:haetae-micro} summarises the HAETAE microbenchmark results together with the public key and signature sizes taken from the parameter sets in Table~\ref{tab:haetae-params}. All timings are given in microseconds (µs).

\begin{table}[t]
\caption{standalone HAETAE microbenchmarks}
\label{tab:haetae-micro}
\centering
\begin{tabular}{lrrrrrr}
\hline
Scheme      & PK [B] & Sig [B] & Keygen [µs] & Sign [µs] & Verify [µs] \\
\hline
HAETAE-120  &  992 & 1474 &  390.4 & 1654.7 &  71.5 \\
HAETAE-180  & 1472 & 2349 &  626.3 & 2083.5 & 132.6 \\
HAETAE-260  & 2080 & 2948 &  539.6 & 2925.0 & 158.4 \\
\hline
\end{tabular}
\end{table}

Overall, the HAETAE variants achieve key and signature sizes that are noticeably smaller than those of ML-DSA at similar security levels, but somewhat larger than the most compact Falcon and Hawk parameter sets. In terms of runtime, key generation stays in the sub-millisecond range, comparable to the ML-DSA variants in Table~\ref{tab:timings}, while signing ranges from roughly 1.65~ms to 2.93~ms and verification from about 0.07~ms to 0.16~ms. These verification costs place HAETAE between the faster Falcon/Hawk variants and the slower ML-DSA-87 configuration on our platform. Because HAETAE is not yet integrated into our blockchain prototype, these numbers should be interpreted purely as cryptographic costs; a full blockchain-level evaluation with HAETAE is left to future implementation.

\subsection{Blockchain-Level Observations}

From a blockchain perspective, two aspects are particularly relevant:

\begin{enumerate}
    \item \textbf{Validation throughput:} For a block containing 1000 transactions, the measured average block validation times range from about 47~ms (Hawk-512) and 52~ms (Falcon-512) up to approximately 234~ms for ML-DSA-87. 
    These values are measured by validating a 1000-transaction block 100 times and dividing the total time by the number of transactions.
    In a high-throughput blockchain, lower verification times translate into higher sustainable transaction rates per validating node \cite{article,10288193}.
    \item \textbf{Bandwidth and storage:} As shown in Table~\ref{tab:sizes}, ML-DSA at higher security levels produces noticeably larger blocks than Falcon or Hawk for the same number of transactions. For example, ML-DSA-87 yields a block of nearly 10~MB for 1000 transactions, whereas Falcon-512 and Hawk-512 stay around 2.5--2.7~MB. For blockchains where block size limits are tight or where network bandwidth is constrained, schemes with smaller signatures and transactions may be advantageous \cite{inbook,cryptoeprint:2022/1151}.
\end{enumerate}

These preliminary results suggest that, at least on our test platform and for the simple prototype considered here, Falcon and Hawk offer attractive trade-offs between performance and data size compared to ML-DSA, especially for verification-heavy workloads. However, ML-DSA benefits from its simpler design and standardisation status, while Hawk remains a non-standard scheme whose long-term security and robustness are still being studied \cite{Ducas_Kiltz_Lepoint_Lyubashevsky_Schwabe_Seiler_Stehlé_2018,huanghttps,cryptoeprint:2022/1151}.

A more comprehensive evaluation would consider additional metrics such as memory usage, energy consumption, and the impact of cryptographic cost on end-to-end block propagation delays in a multi-node network. In future work, we plan to extend our prototype to incorporate HAETAE \cite{cryptoeprint:2023/624} and other emerging schemes, and to perform experiments in a distributed setting to better understand the trade-offs involved in migrating real-world blockchains to post-quantum digital signatures.

\section{Conclusion}

The results in Section~V provide a view of how different post-quantum signature schemes behave in a simple blockchain prototype. In the tests, ML-DSA, Falcon, and Hawk all achieve verification times that are compatible with block validation at the scale of thousands of transactions per block, but they do so with different trade-offs in terms of key sizes, signature sizes, and implementation complexity. ML-DSA offers the simplest design and is standardised by NIST, but has comparatively large public keys, signatures, and block sizes. Falcon and Hawk, achieve significantly smaller signatures and faster verification, which leads to more compact blocks and lower validation latency, but they have more complex constructions.

From a blockchain point of view, two aspects appear particularly important. First, verification cost has a direct impact on validation throughput. For a block with 1000 transactions, Falcon-512 and Hawk-512 validate in tens of milliseconds, whereas ML-DSA-87 reaches a few hundred milliseconds, suggesting that Falcon- and Hawk-based designs could support higher transaction rates per validating node under similar hardware assumptions. Second, signature and key sizes translate almost linearly into transaction and block sizes: ML-DSA-87 produces blocks close to 10~MB in our setup, while Falcon-512 and Hawk-512 stay around 2.5--2.7~MB for the same workload.

On the other hand, HAETAE achieves key and signature sizes smaller than those of ML-DSA at comparable security levels, with verification costs that lie between the Falcon/Hawk variants. Because HAETAE is not yet integrated into the blockchain prototype and is only evaluated via microbenchmarks, these numbers should be interpreted as indicative of its cryptographic cost rather than as full system-level metrics. Overall, the experiments highlight that there is no single scheme that dominates on all areas, instead, practitioners will need to balance implementation complexity, size overhead, and verification latency when selecting a post-quantum signature for blockchain use.

\section{Future Work}

This prototype has several limitations that suggest directions for future work. First, all experiments are conducted on a single-node prototype without networking or consensus, so they do not capture effects such as block propagation delays, peer-to-peer bandwidth constraints, or interactions with block production protocols. Extending the prototype to a multi-node environment, with realistic network latencies and block propagation, would allow us to study how cryptographic costs influence end-to-end throughput.

Second, we currently evaluate only a fixed transaction template (simple transfers between two accounts) and a single block size of 1000 transactions. Future work could explore more diverse workloads, including smart-contract style transactions, varying block sizes, and different mixes of transaction types. 

Third, our HAETAE evaluation is limited to standalone microbenchmarks, the reference implementation depends on GCC-specific features and OpenSSL-based randomness that do not compile cleanly under our MSVC-based Windows environment. Porting HAETAE to our toolchain or moving the blockchain prototype to a Linux-based environment would make it possible to measure HAETAE at the blockchain level using the same methodology as for ML-DSA, Falcon, and Hawk, and to confirm whether its advantages in size and verification time translate into better block-level performance.

Beyond raw speed and size, future development could also include memory usage and energy consumption on different hardware platforms, which will consider nodes running on embedded or energy constrained devices. Together, these changes would provide a more complete picture of the costs and trade-offs involved in migrating current blockchains to quantum-secure digital signatures.

\section*{Acknowledgment}
The author would like to thank Associate Prof. Sedat Akleylek of the Institute of Computer Science,
University of Tartu, for supervision and valuable feedback on this work.

\bibliographystyle{IEEEtran}
\bibliography{refs}

\end{document}